 \documentclass[final,1p,times]{elsarticle}
\usepackage{graphicx}
\usepackage{amssymb}
 \usepackage{amsthm}

\usepackage{lineno}
\usepackage{amsmath}
\newtheorem{theorem}{Theorem}
\newtheorem{proposition}{Proposition}
\newtheorem{remark}{Remark}

\begin{document}
	\begin{center}
	\huge \bf	Phase transition in a power-law uniform hypergraph
	\end{center}
	
	\vspace*{1\baselineskip} 
	
	\begin{center}
		Mingao Yuan\\
			\vspace*{1\baselineskip} 
		\textit{Department of Statistics,}\\
		\textit{North Dakota State University,}\\
		\textit{Fargo, ND 58102, USA}\\
		E-mail: mingao.yuan@ndsu.edu
		
	\end{center}

\begin{center}	
\today	
\end{center}

	\vspace*{1\baselineskip} 
	
\begin{center}	
	\bf Abstract	
\end{center}

We propose a power-law $m$-uniform random hypergraph on $n$ vertexes. In this hypergraph, each vertex is independently assigned a random weight from a power-law distribution with exponent $\alpha\in(0,\infty)$ and the hyperedge probabilities are defined as functions of the random weights. We characterize the number of hyperedge and the number of loose 2-cycle. There is a phase transition phenomenon for the number of hyperedge at $\alpha=1$. Interestingly, for the number of loose 2-cycle, phase transition occurs at both $\alpha=1$ and $\alpha=2$. These results highlights the significent difference between the proposed random hypergraph and the random Erd\"{o}s-R\'{e}nyi hypergraph.


	\section{Introduction}

A graph $\mathcal{G}$ on $n$ nodes consists of a set of nodes $V_n=\{1,2,\dots,n\}$ and an edge set $E$. Each edge in $E$ connects a pair of nodes.
Hypergraph generalizes the concept of graph by allowing a hyperedge to contain more than two nodes. This generalization can accommodate the higher order interaction among nodes in complex network data(\cite{CK10,ABB06,GZCN09,N01,GD17,ER05,B93,ACKZ15}).  
Given an integer $m\geq2$, the \textit{$m$-uniform hypergraph} $\mathcal{H}_m=(V_n, E)$ is defined as follows. Each hyperedge in $E$ is a subset of $V_n$ with exactly $m$ distinct nodes. A hypergraph is usually represented as an adjacency tensor $A\in\{0,1\}^{\tiny \underbrace{n\times n\times\dots\times n}_m}$, where $A_{i_1i_2\dots i_m}=1$ if  $\{i_1,i_2,\dots, i_m\}\in E$ and $A_{i_1i_2\dots i_m}=0$ otherwise.
In this paper, the adjacency tensor $A$ is assumed to be symmetric, that is, $A_{i_1i_2\dots i_m}=A_{j_1j_2\dots j_m}$ if $\{i_1,i_2,\dots,i_m\}=\{j_1,j_2,\dots,j_m\}$. When $m=2$, the hypergraph $\mathcal{H}_2$ is just the usual graph model.

 A graph or hypergraph is said to be random if an edge or hyperedge is included in the graph or hypergraph randomly. The simplest random graph is the well-known Erd\"{o}s-R\'{e}nyi model $\mathcal{G}(n,p)$, where each edge is present in the graph with probability $p$ (\cite{ER60}). The graph $\mathcal{G}(n,p)$ is homogeneous in the sense that each node has the same average degree. In practice, graph data usually presents heterogeneity. To accommodate the heterogeneity, miscellaneous inhomogeneous random graphs have been introduced (\cite{S02,BJR07,CL06,BDM06,CL02}) and one of the fundamental tasks is to understand the local properties of the graphs  (\cite{BJR07,CL06,CL02,BDM06, HD20, HBF14,J07,JLN10}). Among those models, power-law random graphs have attracted a lot of attentions in literature and interesting results were obtained (\cite{BJR07,CL06,CL02,BDM06, HD20, HBF14,J07,JLN10}).

  In this paper, we propose a power-law $m$-uniform random hypergraph. In this hypergraph, each vertex is independently assigned a random weight from a power-law distribution with exponent $\alpha\in(0,\infty)$ and the hyperedge probabilities are defined as functions of the random weights. We denote the hypergraph as $\mathcal{H}_m(n,\alpha)$ and study its local properties. Specifically, we investigate the number of hyeredge and the number of 2-loose cycle, the shortest cycle in $\mathcal{H}_m(n,\alpha)$ with $m\geq3$.
  We find that phase transition phenomenon exists for the number of hyperedge $\alpha=1$ and the number of loose 2-cycle at $\alpha=2$ respectively. The main results are presented in section 2 and the proof is postponed to section 3.

\section{Model and Main Result}
For positive constants $\lambda,\alpha, x_0$, let $W$ be a random variable following some distribution with
\begin{equation}\label{powerlaw}
\mathbb{P}(W>x)=\lambda x^{-\alpha},\ \ x\geq x_0.
\end{equation}
The parameter $\alpha$ controls the existence of moments of the random variable $W$. If $\alpha<1$, the mean does not exist while for $\alpha>1$, the random variable has finite mean. The distribution assumption (\ref{powerlaw}) is common in power law random graphs (\cite{JLN10,BDM06,NR06,JLS19}). 
Given independent and identically distributed random variables $W_1,\dots, W_n$ from a distribution satisfying (\ref{powerlaw}), we define a random $m$-uniform hypergraph $\mathcal{H}_m(n,\alpha)$ as
\[\mathbb{P}(A_{i_1\dots i_m}=1)=\frac{W_{i_1}W_{i_2}\dots W_{i_m}}{n+W_{i_1}W_{i_2}\dots W_{i_m}},\]
where $A_{i_1\dots i_m}\ (1\leq i_1<i_2<\dots<i_m\leq n)$ are assumed to be independent given the random weights $W_i\ (1\leq i\leq n)$. In this paper, $\mathcal{H}_m(n,\alpha)$ is called power-law random $m$-uniform hypergraph. When $m=2$, $\mathcal{H}_2(n,\alpha)$ is just the generalized random graph in \cite{BDM06}.

In the subsequent, we study the local properties of the random hypergraph $\mathcal{H}_m(n,\alpha)$. Firstly, we consider the number of hyperedge, denoted as $\mathcal{E}_m$. The following theorem gives the exact order of $\mathcal{E}_m$.

\begin{theorem}\label{phase1}
For the random hypergraph $\mathcal{H}_m(n,\alpha)$, the following result holds.
\begin{itemize}
    \item If $\alpha>1$, then $\mathcal{E}_m=\left(1+o_p(1)\right)\frac{[\mathbb{E}(W)]^m}{m!}n^{m-1}$ for integer $m\geq2$;
    \item If $\alpha=1$, then $\mathcal{E}_m=\left(1+o_p(1)\right)cn^{m-1}(\log n)^m$ for some constant $c>0$ and $m=2,3,4$;
    \item If $\alpha<1$, then  $\mathcal{E}_m=\left(1+o_p(1)\right)\frac{\pi\alpha^{m}\lambda^m}{\sin (\alpha\pi) m!(m-1)!}n^{m-\alpha}[\log(n)]^{m-1}$, $m=2,3,4$.
\end{itemize}
\end{theorem}
Based on Theorem \ref{phase1},
there is a phase transition phenomenon for the number of hyperedge at $\alpha=1$: for $\alpha>1$, the hypergraph has $\frac{[\mathbb{E}(W)]^m}{m!}n^{m-1}$ hyperedge asymptotically; for $\alpha<1$, the number of hyperedge is  $\frac{\pi\alpha^{m}\lambda^m}{\sin (\alpha\pi) m!(m-1)!}n^{m-\alpha}[\log(n)]^{m-1}$ asymptotically; when $\alpha=1$, the hypergraph has $Cn^{m-1}(\log n)^m$ hyperedges asymptotically. Interestingly, the number of hyperedge decreases as $\alpha$ increases and $\alpha=1$ is a discontinuity point. Besides, the mean of $W$ does not exist if $\alpha\leq 1$ and in this case the hypergraph contains more hyperedges than $\alpha>1$.

Next, we consider the number of short cycles in $\mathcal{H}_m(n,\alpha)$.
Interestingly, a cycle in a hypergraph can be defined in various ways (\cite{JN09,GL12,FO17,HM19, QYW14,KKMO11}) and different types of cycle could contain different information about the global structure of a hypergraph (\cite{YS21}). In this paper, we focus on loose 2-cycle, the shortest cycle in $\mathcal{H}_m(n,\alpha)$ for $m\geq3$. A \textit{loose} 2-cycle in a hypergraph consists of two hyperedges with exactly two common nodes. For instance, the two hyperedges $\{\{1,2,3\},\{1,2,4\}\}$ is a loose 2-cycle in 3-uniform hypergraph. A loose 2-cycle in a graph is trivial since it is just an edge. Non-trivial loose 2-cycle only exists in hypergraph with $m\geq 3$. This differentiates hypergraph ($m\geq3$) from graph ($m=2$).

Below we study the number of loose 2-cycle, denoted as $\mathcal{C}_2$. The following theorem characterizes the order of the expectation of $\mathcal{C}_2$ in $\mathcal{H}_3(n,\alpha)$.
\begin{theorem}\label{phase2}
For $m=3$ and positive constants $c_1,c_2,c_3,c_4$ dependent on the distribution of $W$, the following holds. 
\begin{itemize}
    \item If $\alpha>2$, then $\mathbb{E}(\mathcal{C}_2)=c_1n^2$; moreover, $\mathcal{C}_2=\mathbb{E}(\mathcal{C}_2)(1+o_p(1))$ if $\alpha>3$;
    \item If $\alpha=2$, then $\mathbb{E}(\mathcal{C}_2)=c_2n^2(\log n)^2$;
    \item If $\alpha<2$ and $\alpha\neq 1$, then $\mathbb{E}(\mathcal{C}_2)=c_3n^{4-\alpha}\log n$; moreover, $\mathcal{C}_2=\mathbb{E}(\mathcal{C}_2)(1+o_p(1))$ if $\alpha<1$;
    \item If $\alpha=1$, then $\mathbb{E}(\mathcal{C}_2)=c_4n^3(\log n)^2$; moreover, $\mathcal{C}_2=\mathbb{E}(\mathcal{C}_2)(1+o_p(1))$.
\end{itemize}
\end{theorem}

Based on Theorem \ref{phase2}, a phase transition phenomenon exists for the expected number of 2-loose cycle at $\alpha=1$ and $\alpha=2$. As $\alpha$ increases, the order of $\mathbb{E}(\mathcal{C}_2)$ changes from $n^{4-\alpha}\log n$ to $n^2$ with discontinuity at $\alpha=1$ and $\alpha=2$.

\begin{remark}
	 Theorem \ref{phase1} and Theorem \ref{phase2} shows $\mathcal{H}_3(n,\alpha)$ is significantly different from the corresponding random Erd\"{o}s-R\'{e}nyi hypergraph. To illustrate this, let $\alpha=1$. Based on Theorem \ref{phase1}, $\mathcal{H}_3(n,1)$ has $cn^2(\log n)^3$ hyperedges on average for large $n$. Denote the corresponding random Erd\"{o}s-R\'{e}nyi hypergraph as $\mathcal{G}_3(n,p)$, where hyperedge appears independently with probability $p=\frac{cn^2(\log n)^3}{\binom{n}{3}}$. Then $\mathcal{G}_3(n,p)$ and $\mathcal{H}_3(n,1)$ have the same hyperedge probability and hence the same expected number of hyperedges for large $n$. However, Theorem \ref{phase2} says $\mathcal{H}_3(n,1)$ contains $c_4n^3(\log n)^2$ loose 2-cycles on average while $\mathcal{G}_3(n,p)$ has $cn^2(\log n)^6$ loose 2-cycles on average. Hence $\mathcal{H}_3(n,1)$ differs significantly from $\mathcal{G}_3(n,p)$.
\end{remark}

\section{Proof of Main Result}
We prove Theorem \ref{phase1} in two parts $\alpha>1$ and $\alpha<1$ separately. For convenience, denote
\[p_{i_1\dots i_m}=\frac{W_{i_1}W_{i_2}\dots W_{i_m}}{n+W_{i_1}W_{i_2}\dots W_{i_m}}.\]

\subsection{Proof of the case $\alpha>1$ of Theorem \ref{phase1} .}
We shall employ Markov's inequality to prove the result. Below we find the expectation of $\mathcal{E}_m$ and bound the variance. 

We firstly find the mean hyperedge probability. Using the adjacency matrix $A$, we can express the number of hyperedge as
\[\mathcal{E}_m=\sum_{1\leq i_1<i_2<\dots<i_m\leq n}A_{i_1\dots i_m}.\]
Hence, we have
\begin{eqnarray}\nonumber
\mathbb{E}[\mathcal{E}_m]&=&\sum_{1\leq i_1<i_2<\dots<i_m\leq n}\mathbb{E}\left[p_{i_1\dots i_m}\right]\\ \nonumber
&\leq& \sum_{1\leq i_1<i_2<\dots<i_m\leq n}\mathbb{E}\left[\frac{W_{i_1}W_{i_2}\dots W_{i_m}}{n}\right]\\ \label{eupper}
&=&\frac{n^{m-1}}{m!}[\mathbb{E}(W)]^m(1+o(1)).
\end{eqnarray}
Next, we find a lower bound of $\mathbb{E}[\mathcal{E}_m]$ that matches the upper bound. To this end, define $a_n=n^{\frac{1}{4m}}$, $a\wedge b=\min\{a,b\}$. Then $W_{i_1}\wedge a_n\leq a_n$ and
\[\frac{(W_{i_1}\wedge a_n)(W_{i_2}\wedge a_n)\dots (W_{i_m}\wedge a_n)}{n+(W_{i_1}\wedge a_n)(W_{i_2}\wedge a_n)\dots (W_{i_m}\wedge a_n)}\leq \frac{W_{i_1}W_{i_2}\dots W_{i_m}}{n+W_{i_1}W_{i_2}\dots W_{i_m}},\]
and
\begin{eqnarray}\nonumber 
&&\frac{\left(\sum_{i=1}^nW_i\wedge a_n\right)^m}{n^m}-\frac{1}{n^{m-1}}\sum_{\substack{1\leq i_1,\dots,i_m\leq n\\  
 |\{i_1,\dots,i_m\}|=m}}\frac{W_{i_1}W_{i_2}\dots W_{i_m}}{n+W_{i_1}W_{i_2}\dots W_{i_m}}\\  \nonumber 
 &\leq &\frac{1}{n^m}\sum_{\substack{1\leq i_1,\dots,i_m\leq n\\ |\{i_1,\dots,i_m\}|\leq m-1}}(W_{i_1}\wedge a_n)(W_{i_2}\wedge a_n)\dots (W_{i_m}\wedge a_n)\\ \nonumber 
&&+\frac{1}{n^{m-1}}\sum_{\substack{1\leq i_1,\dots,i_m\leq n\\ |\{i_1,\dots,i_m\}|=m}}\frac{(W_{i_1}\wedge a_n)^2(W_{i_2}\wedge a_n)^2\dots (W_{i_m}\wedge a_n)^2}{n(n+(W_{i_1}\wedge a_n)(W_{i_2}\wedge a_n)\dots (W_{i_m}\wedge a_n))}\\  \nonumber 
&\leq&\frac{n^{m-1}a_n^m}{n^m}+\frac{n^{m}a_n^{2m}}{n^{m+1}}\\ \label{elower}
&=&o(1).
\end{eqnarray}
Note that $\frac{\sum_{i=1}^nW_i\wedge a_n}{n}\rightarrow \mathbb{E}(W)$ in probability. Hence, $\frac{\left(\sum_{i=1}^nW_i\wedge a_n\right)^m}{n^m}\rightarrow [\mathbb{E}(W)]^m$ in probability. By (\ref{elower}), we have
\begin{eqnarray*}
\frac{\left(\sum_{i=1}^nW_i\wedge a_n\right)^m}{n^m}&\leq& \frac{1}{n^{m-1}}\sum_{\substack{1\leq i_1,\dots,i_m\leq n\\  
 |\{i_1,\dots,i_m\}|=m}}p_{i_1\dots i_m}+o(1),\\
 &\leq&\frac{1}{n^{m}}\sum_{\substack{1\leq i_1,\dots,i_m\leq n\\  
 |\{i_1,\dots,i_m\}|=m}}W_{i_1}W_{i_2}\dots W_{i_m}+o(1)
 \end{eqnarray*}
 \begin{eqnarray*}
 &=&\left(\frac{\sum_{i=1}^nW_i}{n}\right)^{m}-\frac{1}{n^{m}}\sum_{\substack{1\leq i_1,\dots,i_m\leq n\\  
 |\{i_1,\dots,i_m\}|\leq m-1}}W_{i_1}W_{i_2}\dots W_{i_m}+o(1).
\end{eqnarray*}
We claim $\frac{1}{n^{m}}\sum_{\substack{1\leq i_1,\dots,i_m\leq n\\  
 |\{i_1,\dots,i_m\}|= k}}W_{i_1}W_{i_2}\dots W_{i_m}=o(1)$ almost surely for $1\leq k\leq m-1$. We prove this by induction.
 According to the Marcinkiewicz-Zygmund strong law of large number, for any integer $t\geq2$, it follows
 \[\frac{1}{n^{t}}\sum_{i=1}^nW_i^t=o(1),\ a.e.\]
For positive integers $t_1,t_2$ such that $t_1+t_2=m$, we have
 \begin{eqnarray*}
\frac{1}{n^{m}}\sum_{i\neq j}^nW_i^{t_1}W_j^{t_2}=\left(\frac{1}{n^{t_1}}\sum_{i=1}^nW_i^{t_1}\right)\left(\frac{1}{n^{t_2}}\sum_{ j=1}^nW_j^{t_2}\right)-\frac{1}{n^{m}}\sum_{i=1}^nW_i^m=o_p(1).
\end{eqnarray*}
Then
\begin{eqnarray*}
\frac{1}{n^{m}}\sum_{\substack{1\leq i_1,\dots,i_m\leq n\\  
 |\{i_1,\dots,i_m\}|= 2}}W_{i_1}W_{i_2}\dots W_{i_m}&=&\sum_{ t_1+t_2=m}\left(\frac{1}{n^{m}}\sum_{i\neq j}^nW_i^{t_1}W_j^{t_2}\right)=o_p(1).
\end{eqnarray*}
 Suppose the result holds for $k$. Next we consider $k+1$. For positive integers $t_1,\dots, t_{k+1}$ with $t_1+t_2+\dots+t_{k+1}=m$, one has 
 \begin{eqnarray*}
&&\frac{1}{n^{m}}\sum_{\substack{1\leq i_1,\dots,i_m\leq n\\  
 |\{i_1,\dots,i_m\}|= k+1}}W_{i_1}W_{i_2}\dots W_{i_m}\\
 &=&\sum_{ t_1+\dots+t_{k+1}=m}\left(\frac{1}{n^{m}}\sum_{|\{i_1,\dots,i_{k+1}\}|=k+1}^nW_{i_1}^{t_1}\dots W_{i_{k+1}}^{t_{k+1}}\right)\\
 &=&\sum_{ t_1+\dots+t_{k+1}=m}\left(\frac{1}{n^{t_1}}\sum_{i=1}^nW_i^{t_1}\right)\dots\left(\frac{1}{n^{t_{k+1}}}\sum_{ j=1}^nW_j^{t_{k+1}}\right)\\
 &&-\sum_{ t_1+\dots+t_{k+1}=m}\left(\frac{1}{n^{m}}\sum_{|\{i_1,\dots,i_{k+1}\}|\leq k}^nW_{i_1}^{t_1}\dots W_{i_{k+1}}^{t_{k+1}}\right)\\
 &=&o_p(1).
\end{eqnarray*}

 Hence, by the generalized Lebesgue dominated convergence theorem, we have \[\frac{m!}{n^{m-1}}\mathbb{E}[\mathcal{E}_m]\geq \mathbb{E}\left[\frac{\left(\sum_{i=1}^nW_i\wedge a_n\right)^m}{n^m}\right]=[\mathbb{E}(W)]^m(1+o(1)).\]
Consequently, $\mathbb{E}[\mathcal{E}_m]=\frac{n^{m-1}}{m!}[\mathbb{E}(W)]^m(1+o(1))$.

Next we bound the variance of $\mathcal{E}_m$. 
It is clear that $n\mathbb{E}(p_{i_1\dots i_m})=O(1)$.
By the definition of variance, one has
\begin{eqnarray*}
Var(\mathcal{E}_m)&=&\mathbb{E}(\mathcal{E}_m^2)-[\mathbb{E}(\mathcal{E}_m)]^2\\
&=&\mathbb{E}\left[\sum_{\substack{1\leq i_1<i_2<\dots<i_m\leq n\\ 1\leq j_1<j_2<\dots<j_m\leq n}}A_{i_1\dots i_m}A_{j_1\dots j_m}\right]-\sum_{\substack{1\leq i_1<i_2<\dots<i_m\leq n\\ 1\leq j_1<j_2<\dots<j_m\leq n}}\mathbb{E}[p_{i_1\dots i_m}]\mathbb{E}[p_{j_1\dots j_m}]
 \end{eqnarray*}
 \begin{eqnarray*}
&=&\mathbb{E}\left[\sum_{\substack{1\leq i_1<i_2<\dots<i_m\leq n\\ 1\leq j_1<j_2<\dots<j_m\leq n\\
|\{i_1,\dots,i_m\}\cap\{j_1,\dots,j_m\}|=0}}A_{i_1\dots i_m}A_{j_1\dots j_m}\right]+\mathbb{E}\left[\sum_{\substack{1\leq i_1<i_2<\dots<i_m\leq n\\ 1\leq j_1<j_2<\dots<j_m\leq n\\
1\leq |\{i_1,\dots,i_m\}\cap\{j_1,\dots,j_m\}|\leq m-1}}A_{i_1\dots i_m}A_{j_1\dots j_m}\right]\\
&&+\mathbb{E}(\mathcal{E}_m) -\sum_{\substack{1\leq i_1<i_2<\dots<i_m\leq n\\ 1\leq j_1<j_2<\dots<j_m\leq n}}\mathbb{E}[p_{i_1\dots i_m}]\mathbb{E}[p_{j_1\dots j_m}]\\
&=& \sum_{\substack{1\leq i_1<i_2<\dots<i_m\leq n\\ 1\leq j_1<j_2<\dots<j_m\leq n\\
|\{i_1,\dots,i_m\}\cap\{j_1,\dots,j_m\}|=0}}\mathbb{E}[A_{i_1\dots i_m}]\mathbb{E}[A_{j_1\dots j_m}] +\sum_{\substack{1\leq i_1<i_2<\dots<i_m\leq n\\ 1\leq j_1<j_2<\dots<j_m\leq n\\
1\leq |\{i_1,\dots,i_m\}\cap\{j_1,\dots,j_m\}|\leq m-1}}\mathbb{E}[A_{i_1\dots i_m}A_{j_1\dots j_m}]\\
&&+\mathbb{E}(\mathcal{E}_m) -\sum_{\substack{1\leq i_1<i_2<\dots<i_m\leq n\\ 1\leq j_1<j_2<\dots<j_m\leq n}}\mathbb{E}[p_{i_1\dots i_m}]\mathbb{E}[p_{j_1\dots j_m}]\\
&\leq &\mathbb{E}(\mathcal{E}_m)+\sum_{\substack{1\leq i_1<i_2<\dots<i_m\leq n\\ 1\leq j_1<j_2<\dots<j_m\leq n\\
1\leq |\{i_1,\dots,i_m\}\cap\{j_1,\dots,j_m\}|\leq m-1}}\mathbb{E}[p_{i_1\dots i_m}p_{j_1\dots j_m}].
\end{eqnarray*}
Note that $p_{i_1i_2\dots i_m}\leq1$.
For $2\leq t\leq m-1$, we have
\[\sum_{\substack{1\leq i_1<i_2<\dots<i_m\leq n\\ 1\leq j_1<j_2<\dots<j_m\leq n\\
|\{i_1,\dots,i_m\}\cap\{j_1,\dots,j_m\}|=t}}p_{i_1\dots i_m}p_{j_1\dots j_m}\leq n^{2m-t}\leq n^{2m-2}.\]
Suppose $i_1=j_1,\dots, i_t=j_t$. Then
\begin{eqnarray*}
&&\sum_{\substack{1\leq i_1<\dots<i_t<i_{t+1}<\dots<i_m\leq n\\ 1\leq i_1< \dots<i_t<j_{t+1}<\dots<j_m\leq n\\
|\{i_{t+1},\dots,i_m\}\cap\{j_{t+1},\dots,j_m\}|=0}}p_{i_1\dots i_m}p_{j_1\dots j_m}\\
&\leq&\frac{(\sum_{i_1=1}^nW_{i_1}^2)\dots (\sum_{i_t=1}^nW_{i_t}^2)(\sum_{i_{t+1}=1}^nW_{i_{t+1}})\dots (\sum_{i_m=1}^nW_{i_m})(\sum_{j_{t+1}=1}^nW_{j_{t+1}})\dots (\sum_{j_m=1}^nW_{j_m})}{n^2}\\
&\leq&n^{2m-2}\left(\frac{\sum_{i_1=1}^nW_{i_1}^2}{n^2}\right)\dots \left(\frac{\sum_{i_t=1}^nW_{i_t}^2}{n^2}\right)\left(\frac{\sum_{i_{t+1}}^nW_{i_{t+1}}}{n}\right)\dots \left(\frac{\sum_{i_m}^nW_{i_m}}{n}\right) \left(\frac{\sum_{j_{t+1}}^nW_{j_{t+1}}}{n}\right)\dots \left(\frac{\sum_{j_m}^nW_{j_m}}{n}\right)\\
&=&o_p(n^{2m-2}),
\end{eqnarray*}
where we used the fact that $\frac{\sum_{i_1=1}^nW_{i_1}^2}{n^2}=o_p(1)$ by the Marcinkiewicz-Zygmund strong law of large number and $\frac{\sum_{i_m}^nW_{i_m}}{n}=O_p(1)$. By the dominated convergence theorem, for $2\leq t\leq m-1$, we have
\[\frac{1}{n^{2m-2}}\sum_{\substack{1\leq i_1<i_2<\dots<i_m\leq n\\ 1\leq j_1<j_2<\dots<j_m\leq n\\
t\leq |\{i_1,\dots,i_m\}\cap\{j_1,\dots,j_m\}|\leq m-1}}\mathbb{E}[p_{i_1\dots i_m}p_{j_1\dots j_m}]=o(1).\]

For $|\{i_1,\dots,i_m\}\cap\{j_1,\dots,j_m\}|=1$, without loss of generality, suppose that $i_1=j_1$ and $|\{i_2,\dots,i_m\}\cap\{j_2,\dots,j_m\}|=0$. Let $\beta$ be a constant such that $1<\beta<\min\{2,\alpha\}$. Then we have
\begin{eqnarray*}
\mathbb{E}[W_1^{\beta}]&=&\int_0^{x_0^{\beta}}\mathbb{P}(W_1> t^{\frac{1}{\beta}})dt+\int_{x_0^{\beta}}^{\infty}\mathbb{P}(W_1> t^{\frac{1}{\beta}})dt\\
&\leq&x_0^{\beta}+\int_{x_0^{\beta}}^{\infty}\lambda t^{-\frac{\alpha}{\beta}}dt\\
&=&x_0^{\beta}+\frac{\lambda}{\frac{\alpha}{\beta}-1}x_0^{\frac{\alpha}{\beta}-1}.
\end{eqnarray*}
Hence $\mathbb{E}[W_1^{\beta}]=O(1)$ and 
\[\mathbb{E}[p_{i_1i_2\dots i_m}^{\beta}]\leq \mathbb{E}\left[\frac{(W_{i_1}\dots W_{i_m})^{\beta}}{n^{\beta}}\right]=\frac{1}{n^{\beta}}O(1).\]
By the Cauchy-Schwarz inequality and the fact that $0\leq  p_{i_1\dots i_m}\leq1$, we have 
\begin{eqnarray*}
\sum_{\substack{1\leq i_1<i_2<\dots<i_m\leq n\\ 1\leq i_1<j_2<\dots<j_m\leq n\\
|\{i_2,\dots,i_m\}\cap\{j_2,\dots,j_m\}|=1}}\mathbb{E}[p_{i_1\dots i_m}p_{j_1\dots j_m}]&\leq& \sum_{\substack{1\leq i_1<i_2<\dots<i_m\leq n\\ 1\leq i_1<j_2<\dots<j_m\leq n\\
|\{i_2,\dots,i_m\}\cap\{j_2,\dots,j_m\}|=1}}\sqrt{\mathbb{E}[p_{i_1i_2\dots i_m}^2]\mathbb{E}[p_{i_1j_2\dots j_m}^2]}\\
&\leq&\sum_{\substack{1\leq i_1<i_2<\dots<i_m\leq n\\ 1\leq i_1<j_2<\dots<j_m\leq n\\
|\{i_2,\dots,i_m\}\cap\{j_2,\dots,j_m\}|=1}}\sqrt{\mathbb{E}[p_{i_1i_2\dots i_m}^{\beta}]\mathbb{E}[p_{i_1j_2\dots j_m}^{\beta}]}\\
&=&n^{2m-1-\beta}O(1)
\end{eqnarray*}
Hence, we have
\[\frac{1}{n^{2m-2}}\sum_{\substack{1\leq i_1<i_2<\dots<i_m\leq n\\ 1\leq i_1<j_2<\dots<j_m\leq n\\
|\{i_2,\dots,i_m\}\cap\{j_2,\dots,j_m\}|=0}}\mathbb{E}[p_{i_1\dots i_m}p_{j_1\dots j_m}]=O\left(\frac{1}{n^{\beta-1}}\right)=o(1). \]

Now the variance of $\mathcal{E}_m$ can be bounded as follows.
\[Var(\mathcal{E}_m)\leq\mathbb{E}(\mathcal{E}_m)+o(n^{2m-2}).\]
For any constant $\epsilon>0$, by Markov's inequality, one gets
\[\mathbb{P}\left(\frac{m!|\mathcal{E}_m-\mathbb{E}(\mathcal{E}_m)|}{n^{m-1}}>\epsilon\right)\leq\frac{m!^2Var(\mathcal{E}_m)}{n^{2(m-1)}\epsilon^2}=o(1).\]
Then the proof for the case $\alpha>1$ is complete.

\subsection{Proof of the case $\alpha<1$ of Theorem \ref{phase1}.}

To prove the case $\alpha<1$ in Theorem \ref{phase1}, we prove a more general result as below. Let
\[p_{i_1\dots i_m}=\frac{W_{i_1}W_{i_2}\dots W_{i_m}}{n^{\tau}+W_{i_1}W_{i_2}\dots W_{i_m}}.\]
\begin{proposition}\label{edge}
Let $\alpha<1$ and $m\in\{2,3,4\}$. If $ \tau\leq\frac{1}{\alpha}$, then 
\[\mathcal{E}_m=\left(1+o_p(1)\right)\frac{\pi\alpha^{m}\tau^{m-1}\lambda^m}{\sin (\alpha\pi)(m-1)!}\frac{n^{(m)}}{n^{\alpha\tau}}[\log(n)]^{m-1}.\]
\end{proposition}

\textit{Proof of Proposition \ref{edge}.}  
For large $x>x_0^m$, firstly we evaluate the probability $\mathbb{P}\left(W_{1}\dots W_{m}>x\right)$ with $m=2,3,4$. We claim for large $x$
\begin{equation}\label{ind0}
\mathbb{P}\left(W_{1}\dots W_{m}>x\right)=\frac{\lambda^m\alpha^{m-1}x^{-\alpha}\left(\log x\right)^{m-1}}{(m-1)!}(1+o(1)),\ \ m=2,3,4.
\end{equation}
Let $f(w)$ be the probability density function of $W$. For $x> x_0$, it is easy to get $f(x)=\lambda\alpha x^{-\alpha-1}$.
Let $x> x_0^m$ and $x$ be large below. Note that
\begin{eqnarray}\nonumber
\mathbb{P}\left(W_{1} W_{2}>x\right)&=&\mathbb{P}\left(W_{1}>\frac{x}{W_{2}},W_2<\frac{x}{x_0}\right)
+\mathbb{P}\left(W_{1}>\frac{x}{W_{2}},W_2\geq\frac{x}{x_0}\right)\\ \nonumber
&=&\int_0^{x_0}f(w_2)\lambda x^{-\alpha}w_2^{\alpha}dw_2+\lambda^2\alpha x^{-\alpha}\int_{x_0}^{\frac{x}{x_0}}w_2^{-1}dw_2    +O\left(\lambda x^{-\alpha}\right) 
 \end{eqnarray}
 \begin{eqnarray}\nonumber
&=&\lambda^2\alpha x^{-\alpha}\int_{x_0}^{\frac{x}{x_0}}w_2^{-1}dw_2+O\left(\lambda x^{-\alpha}\right)\\ \label{induction1}
&=&\lambda^2\alpha x^{-\alpha}\log (x)(1+o(1)),
\end{eqnarray}
and by (\ref{induction1}) and integration by parts, it follows that
\begin{eqnarray}\nonumber
&&\mathbb{P}\left(W_{1} W_{2}W_{3}>x\right)\\  \nonumber
&=&\mathbb{P}\left(W_{1}>\frac{x}{W_{2}W_{3}},W_2W_{3}<\frac{x}{x_0}\right)
+\mathbb{P}\left(W_{1}>\frac{x}{W_{2}W_{3}},W_2W_{3}\geq\frac{x}{x_0}\right)\\ \nonumber
&=&\lambda x^{-\alpha}\iint_{W_2W_{3}<\frac{x}{x_0}}f(w_2)f(w_3)w_2^{\alpha} w_3^{\alpha}dw_2dw_3+O\left(x^{-\alpha}\log (x)\right)   \\ \nonumber
&=&\lambda x^{-\alpha}\int_{0}^{x_0}\int_{x_0}^{\frac{x}{x_0w_2}}f(w_2)f(w_3)w_2^{\alpha} w_3^{\alpha}dw_3dw_2+\lambda x^{-\alpha}\int_{0}^{x_0}\int_{x_0}^{\frac{x}{x_0w_3}}f(w_2)f(w_3)w_2^{\alpha} w_3^{\alpha}dw_2dw_3\\ \nonumber
&&+\lambda x^{-\alpha}\int_{0}^{x_0}\int_{0}^{x_0}f(w_2)f(w_3)w_2^{\alpha} w_3^{\alpha}dw_3dw_2\\ \nonumber
&&+\lambda x^{-\alpha}\int_{x_0}^{\frac{x}{x_0}}\int_{x_0}^{\frac{x}{x_0w_2}}f(w_2)f(w_3)w_2^{\alpha} w_3^{\alpha}dw_3dw_2+O\left(x^{-\alpha}\log (x)\right)   \\ \nonumber
&=&\lambda^3\alpha^2 x^{-\alpha}\int_{x_0}^{\frac{x}{x_0}}\int_{x_0}^{\frac{x}{x_0w_2}}w_2^{-1}w_3^{-1}dw_3dw_2+O\left(x^{-\alpha}\log (x)\right) 
\end{eqnarray}
\begin{eqnarray}\nonumber
&=&\lambda^3\alpha^2 x^{-\alpha}\int_{x_0}^{\frac{x}{x_0}}\frac{\log x-\log w_2}{w_2}dw_2(1+o(1))+O\left(x^{-\alpha}\log (x)\right)\\ \nonumber
&=&\lambda^3\alpha^2 x^{-\alpha}\left((\log x)^2-\int_{x_0}^{\frac{x}{x_0}}(\log w_2) d(\log w_2)\right)(1+o(1))+O\left(x^{-\alpha}\log (x)\right)\\ \label{induction2}
&=&\frac{\lambda^3\alpha^2 x^{-\alpha}(\log x)^2}{2}(1+o(1))
\end{eqnarray}
For $m=4$, we have
\begin{eqnarray*}
\mathbb{P}\left(W_{1}\dots W_{4}>x\right)&=&\mathbb{P}\left(W_{4}>\frac{x}{W_{1}W_{2}W_{3}},W_1W_{2} W_{3}<\frac{x}{x_0}\right)\\
&&+\mathbb{P}\left(W_{4}>\frac{x}{W_{1}W_{2} W_{3}},W_1W_{2} W_{3}\geq\frac{x}{x_0}\right)\\
&=&\mathbb{P}\left(W_{4}>\frac{x}{W_{1}W_{2} W_{3}},W_1W_{2} W_{3}<\frac{x}{x_0}\right)+O\left(x^{-\alpha}\left(\log x\right)^{2}\right).
\end{eqnarray*}
A similar calculation as in (\ref{induction2}) yields
\begin{eqnarray*}
&&\mathbb{P}\left(W_{4}>\frac{x}{W_{1}W_{2} W_{3}},W_1W_{2} W_{3}<\frac{x}{x_0}\right)\\
&=&(1+o(1))\lambda^4\alpha^3 x^{-\alpha}\int_{x_0}^{\frac{x}{x_0}}\frac{1}{w_3}dw_3\int_{x_0}^{\frac{x}{w_3}}\frac{1}{w_2}dw_2\int_{x_0}^{\frac{x}{w_2w_3}}\frac{1}{w_1}dw_1\\
&=&(1+o(1))\lambda^4\alpha^3 x^{-\alpha}\int_{x_0}^{\frac{x}{x_0}}\frac{1}{w_3}dw_3\int_{x_0}^{\frac{x}{w_3}}\frac{\log x-\log w_2-\log w_3}{w_2}dw_2\\
&=&(1+o(1))\lambda^4\alpha^3 x^{-\alpha}\int_{x_0}^{\frac{x}{x_0}}\frac{1}{w_3}\left((\log x)^2-\log x\log w_3-\frac{1}{2}(\log x-\log w_3)^2-(\log \frac{x}{w_3})\log w_3\right)dw_3\\
&=&(1+o(1))\lambda^4\alpha^3 x^{-\alpha}\frac{1}{2}\int_{x_0}^{\frac{x}{x_0}}\frac{(\log w)^2}{w}dw
 \end{eqnarray*}
 \begin{eqnarray*}
&=&(1+o(1))\lambda^4\alpha^3 x^{-\alpha}\left(\frac{1}{2}(\log x)^3-\int_{x_0}^{\frac{x}{x_0}}\frac{(\log w)^2}{w}dw\right).
\end{eqnarray*}
Hence 
\[\mathbb{P}\left(W_{1}\dots W_{4}>x\right)=\frac{\lambda^4\alpha^3 x^{-\alpha}(\log x)^3}{6}(1+o(1)).\]
By the fact that $\alpha<1$, the hyperedge probability is
\begin{eqnarray}\nonumber
\mathbb{E}[A_{i_1\dots i_m}]&=&\int_0^1\mathbb{P}\left(\frac{W_{i_1}W_{i_2}\dots W_{i_m}}{n^{\tau}+W_{i_1}W_{i_2}\dots W_{i_m}}>t\right)dt\\ \nonumber
&=&\int_0^1\mathbb{P}\left(W_{i_1}\dots W_{i_m}>\frac{n^{\tau}t}{1-t}\right)dt\\ \nonumber
&=&\int_0^{\infty}\mathbb{P}\left(W_{i_1}\dots W_{i_m}>t\right)\frac{n^{\tau}}{(n^{\tau}+t)^2}dt\\ \nonumber
&=&(1+o(1))\int_{x_0^m}^{\infty}\frac{\lambda^m\alpha^{m-1} t^{-\alpha}[\log(t)]^{m-1}}{(m-1)!}\frac{n^{\tau}}{(n^{\tau}+t)^2}dt+\int_0^{x_0^m}\mathbb{P}\left(W_{i_1}\dots W_{i_m}>t\right)\frac{n^{\tau}}{(n^{\tau}+t)^2}dt\\ \nonumber
&=&(1+o(1))\frac{\lambda^m\alpha^{m-1}}{n^{\tau\alpha}(m-1)!}\int_{\frac{x_0^2}{n^{\tau}}}^{\infty}\frac{(\log n^{\tau}+\log x )^{m-1}}{x^{\alpha}(1+x)^2}dx+O\left(\frac{1}{n^{\tau}}\right)\\ \label{esqure}
&=&(1+o(1))\frac{\lambda^m\alpha^{m-1}\tau^{m-1}[\log n]^{m-1}}{n^{\tau\alpha}(m-1)!}\frac{\alpha\pi}{\sin(\alpha\pi)}+O\left(\frac{1}{n^{\tau}}\right),
\end{eqnarray}
where in the last equality we used the fact that 
\[\int_{\frac{x_0^2}{n^{\tau}}}^{\infty}\frac{(\log n^{\tau}+\log x )^{m-1}}{x^{\alpha}(1+x)^2}dx= \frac{\alpha\pi}{\sin(\alpha\pi)}\tau^{m-1}[\log n]^{m-1}(1+o(1)).\]
To see this, note that
\begin{eqnarray*}
&&\int_{\frac{x_0^2}{n^{\tau}}}^{\infty}\frac{(\log n^{\tau}+\log x )^{m-1}}{x^{\alpha}(1+x)^2}dx\\
&=&\int_{\frac{x_0^2}{n^{\tau}}}^{1}\frac{(\log n^{\tau}+\log x )^{m-1}}{x^{\alpha}(1+x)^2}dx+\int_{1}^{\infty}\frac{(\log n^{\tau}+\log x )^{m-1}}{x^{\alpha}(1+x)^2}dx\\
&=&\left(\log n^{\tau}\right)^{m-1}\int_{\frac{x_0^2}{n^{\tau}}}^{1}\frac{1}{x^{\alpha}(1+x)^2}dx+\int_{\frac{x_0^2}{n^{\tau}}}^{1}\frac{(m-1)\left(\log n^{\tau}\right)^{m-2}\log x+\dots +\left(\log x\right)^{m-1}}{x^{\alpha}(1+x)^2}dx\\
&&+\left(\log n^{\tau}\right)^{m-1}\int_{1}^{\infty}\frac{1}{x^{\alpha}(1+x)^2}dx+\int_{1}^{\infty}\frac{(m-1)\left(\log n^{\tau}\right)^{m-2}\log x+\dots +\left(\log x\right)^{m-1}}{x^{\alpha}(1+x)^2}dx\\
\end{eqnarray*}
\begin{eqnarray*}
&=&\left(\log n^{\tau}\right)^{m-1}\int_{\frac{x_0^2}{n^{\tau}}}^{\infty}\frac{1}{x^{\alpha}(1+x)^2}dx+\int_{\frac{x_0^2}{n^{\tau}}}^{1}\frac{(m-1)\left(\log n^{\tau}\right)^{m-2}\log x+\dots +\left(\log x\right)^{m-1}}{x^{\alpha}(1+x)^2}dx+O\left(\left(\log n^{\tau}\right)^{m-2}\right).
\end{eqnarray*}
For $1\leq k\leq m-2$, we have
\begin{eqnarray*}
\left|\int_{\frac{x_0^2}{n^{\tau}}}^{1}\frac{\left(\log x\right)^{k}}{x^{\alpha}(1+x)^2}dx \right|\leq \left(\log n^{\tau}\right)^{k}\int_{\frac{x_0^2}{n^{\tau}}}^{1}\frac{1}{x^{\alpha}(1+x)^2}dx \leq \left(\log n^{\tau}\right)^{k}\int_{\frac{x_0^2}{n^{\tau}}}^{1}x^{-\alpha}dx\leq \frac{\left(\log n^{\tau}\right)^{k}}{1-\alpha}.
\end{eqnarray*}
If $m-1$ is even, then 
\begin{eqnarray*}
0\leq \int_{\frac{x_0^2}{n^{\tau}}}^{1}\frac{\left(\log x\right)^{m-1}}{x^{\alpha}(1+x)^2}dx&\leq& -\frac{1}{1-\alpha}\left(\frac{x_0^2}{n^{\tau}}\right)^{1-\alpha}\left(\log \frac{x_0^2}{n^{\tau}}\right)^{m-1}-\frac{m-1}{1-\alpha}\int_{\frac{x_0^2}{n^{\tau}}}^{1}x^{-\alpha}\left(\log x\right)^{m-2}dx\\
&=&O\left(\left(\log n\right)^{m-2}\right).
\end{eqnarray*}
If $m-1$ is odd, then
\begin{eqnarray*}
0\geq \int_{\frac{x_0^2}{n^{\tau}}}^{1}\frac{\left(\log x\right)^{m-1}}{x^{\alpha}(1+x)^2}dx&\geq& -\frac{1}{1-\alpha}\left(\frac{x_0^2}{n^{\tau}}\right)^{1-\alpha}\left(\log \frac{x_0^2}{n^{\tau}}\right)^{m-1}-\frac{m-1}{1-\alpha}\int_{\frac{x_0^2}{n^{\tau}}}^{1}x^{-\alpha}\left(\log x\right)^{m-2}dx\\
&=&-O\left(\left(\log n\right)^{m-2}\right).
\end{eqnarray*}
Hence,
\[\int_{\frac{x_0^2}{n^{\tau}}}^{\infty}\frac{(\log n^{\tau}+\log x )^{m-1}}{x^{\alpha}(1+x)^2}dx=(1+o(1))\left(\log n^{\tau}\right)^{m-1}\int_{\frac{x_0^2}{n^{\tau}}}^{\infty}\frac{1}{x^{\alpha}(1+x)^2}dx=\frac{\alpha\pi}{\sin(\alpha\pi)}\tau^{m-1}[\log n]^{m-1}(1+o(1)).\]

Next, we bound the variance of $\mathcal{E}_m$. By the proof of case $\alpha>1$, we have
\begin{eqnarray*}
Var(\mathcal{E}_m)
&\leq& \mathbb{E}(\mathcal{E}_m)+\sum_{\substack{1\leq i_1<i_2<\dots<i_m\leq n\\ 1\leq j_1<j_2<\dots<j_m\leq n\\
1\leq |\{i_1,\dots,i_m\}\cap\{j_1,\dots,j_m\}|\leq m-1}}\mathbb{E}\left[\frac{W_{i_1}W_{i_2}\dots W_{i_m}}{n^{\tau}+W_{i_1}W_{i_2}\dots W_{i_m}}\frac{W_{j_1}W_{j_2}\dots W_{j_m}}{n^{\tau}+W_{j_1}W_{j_2}\dots W_{j_m}}\right]\\
&\leq&\mathbb{E}(\mathcal{E}_m)+\sum_{\substack{1\leq i_1<i_2<\dots<i_m\leq n\\ 1\leq j_1<j_2<\dots<j_m\leq n\\
1\leq |\{i_1,\dots,i_m\}\cap\{j_1,\dots,j_m\}|\leq m-1}}\left(\mathbb{E}\left[\frac{W_{i_1}W_{i_2}\dots W_{i_m}}{n^{\tau}+W_{i_1}W_{i_2}\dots W_{i_m}}\right]^2+\mathbb{E}\left[\frac{W_{j_1}W_{j_2}\dots W_{j_m}}{n^{\tau}+W_{j_1}W_{j_2}\dots W_{j_m}}\right]^2\right).
\end{eqnarray*}
For some generic constant $C>0$, simple calculation yields
\begin{eqnarray}\nonumber
\mathbb{E}\left[\frac{W_{i_1}W_{i_2}\dots W_{i_m}}{n^{\tau}+W_{i_1}W_{i_2}\dots W_{i_m}}\right]^2&=&\int_0^{\infty}\frac{2xn^{\tau}}{(n^{\tau}+x)^3}\mathbb{P}\left(W_{i_1}\dots W_{i_m}>x\right)dx\\ \nonumber
&\leq&C\int_{x_0^2}^{\infty}\frac{2xn^{\tau}[\log(x)]^{m-1}}{(n^{\tau}+x)^3x^{\alpha}}dx+O\left(\frac{1}{n^{2\tau}}\right)\\ \nonumber
&=&C\frac{1}{n^{\tau\alpha}}\int_{\frac{x_0^2}{n^{\tau}}}^{\infty}\frac{x[\log(x)+\tau\log n]^{m-1}}{(1+x)^3x^{\alpha}}dx+O\left(\frac{1}{n^{2\tau}}\right)\\ \nonumber
&=&C\frac{[\log(n)]^{m-1}}{n^{\tau\alpha}}+O\left(\frac{1}{n^{2\tau}}\right).
\end{eqnarray}
Hence, $Var(\mathcal{E}_m)=O\left(\mathbb{E}(\mathcal{E}_m)+n^{2m-1-\tau\alpha}[\log(n)]^{m-1}\right)$. Then the desired result follows from the Markov's inequality.

\subsection{Proof of the case $\alpha=1$  of Theorem \ref{phase1}.}
For $\alpha=1$, we have
\begin{eqnarray*}\nonumber
\mathbb{E}[A_{i_1\dots i_m}]&=&\int_0^1\mathbb{P}\left(\frac{W_{i_1}W_{i_2}\dots W_{i_m}}{n+W_{i_1}W_{i_2}\dots W_{i_m}}>t\right)dt\\ \nonumber
&=&(1+o(1))\frac{\lambda^m}{n(m-1)!}\int_{\frac{x_0^2}{n}}^{\infty}\frac{(\log n+\log x )^{m-1}}{x(1+x)^2}dx+O\left(\frac{1}{n}\right)\\ \nonumber
&=&(1+o(1))\frac{\lambda^m}{n(m-1)!}\int_{\frac{x_0^2}{n}}^{1}\frac{(\log n+\log x )^{m-1}}{x(1+x)^2}dx+O\left(\frac{(\log n)^{m-1}}{n}\right).  
\end{eqnarray*}
Note that
\[\frac{1}{4}\int_{\frac{x_0^2}{n}}^{1}\frac{(\log n+\log x )^{m-1}}{x}dx\leq \int_{\frac{x_0^2}{n}}^{1}\frac{(\log n+\log x )^{m-1}}{x(1+x)^2}dx\leq \int_{\frac{x_0^2}{n}}^{1}\frac{(\log n+\log x )^{m-1}}{x}dx,\]
and
\[\int_{\frac{x_0^2}{n}}^{1}\frac{(\log n+\log x )^{m-1}}{x}dx=\int_{\log\frac{x_0^2}{n}}^{0}(\log n+x )^{m-1}dx=\frac{(\log n)^m}{m}(1+o(1)).\]
Hence, $\mathbb{E}(\mathcal{E}_m)=Cn^{m-1}(\log n)^m$. Then the desired result follows by Markov's inequality.

\subsection{Proof of the case $\alpha<2$ and $\alpha\neq1$ of Theorem \ref{phase2} .}
We use Markov's inequality to prove the result.
We firstly bound the probability of 2-loose cycle. Using the adjacency matrix $A$, we can express the number of 2-loose cycle as
\[\mathcal{C}_2=\sum_{1\leq i<j<k<l\leq n}\left(A_{ijk}A_{ijl}+A_{ilk}A_{ijl}+A_{ijk}A_{ikl}+A_{ijk}A_{kjl}+A_{ljk}A_{ijl}+A_{ljk}A_{ikl}\right).\]
Hence, $\mathbb{E}(\mathcal{C}_2)=6n^{(4)}\mathbb{E}(p_{123}p_{143})$.
Next, we find the order of $\mathbb{E}(p_{123}p_{143})$. Note that
\begin{eqnarray}\label{epp1}
\mathbb{E}(p_{123}p_{143})&=&\mathbb{E}(p_{123}p_{143}I[W_1W_3\leq n])+\mathbb{E}(p_{123}p_{143}I[W_1W_3> n]).
\end{eqnarray}
By (\ref{induction1}), it is clear that 
\[0\leq \mathbb{E}(p_{123}p_{143}I[W_1W_3> n])\leq \mathbb{P}(W_1W_3> n)\leq \lambda^2\alpha\frac{\log n}{n^{\alpha}}.\]
It suffices to prove $\mathbb{E}(p_{123}p_{143}I[W_1W_3\leq n])\asymp  \frac{\log n}{n^{\alpha}}$.

Let $f(w)$ be the probability density function of $W$. For $x> x_0$, it is easy to get $f(x)=\lambda\alpha x^{-\alpha-1}$. Let $\Omega=\{(w_1,w_2,w_3,w_4)\in\mathbb{R}^4:w_1w_3\leq n,w_2>0,w_4>0\}$. Then
\begin{eqnarray}\nonumber
&&\mathbb{E}(p_{123}p_{143}I[W_1W_3\leq n])\\ \nonumber
&=&\iiiint_{\Omega}f(w_1)f(w_2)f(w_3)f(w_4)p_{123}p_{143}dw_1dw_2dw_3dw_4\\ \label{epp2}
&=&\iint_{w_1w_3\leq n}f(w_1)f(w_3)\mathbb{E}(p_{123}|W_1=w_1,W_3=w_3)\mathbb{E}(p_{143}|W_1=w_1,W_3=w_3)dw_1dw_2.
\end{eqnarray}
Note that
\begin{eqnarray}\nonumber
&&\mathbb{E}(p_{123}|W_1=w_1,W_3=w_3)\\ \nonumber
&=&\int_0^1\mathbb{P}(p_{123}>t|W_1=w_1,W_3=w_3)\\  \nonumber
&=&\int_0^1\mathbb{P}\left(W_2>\frac{nt}{w_1w_3(1-t)}\right)dt\\  \nonumber
&=&\int_0^{\infty}\mathbb{P}\left(W_2>t\right)\frac{nw_1w_3}{(n+w_1w_3t)^2}dt\\  \nonumber
&=&\int_0^{\infty}\mathbb{P}\left(W_2>\frac{nt}{w_1w_3}\right)\frac{1}{(1+t)^2}dt\\ \nonumber
&=&\int_0^{\frac{w_1w_3x_0}{n}}\mathbb{P}\left(W_2>\frac{nt}{w_1w_3}\right)\frac{1}{(1+t)^2}dt+\int_{\frac{w_1w_3x_0}{n}}^{\infty}\mathbb{P}\left(W_2>\frac{nt}{w_1w_3}\right)\frac{1}{(1+t)^2}dt\\  \nonumber
&\leq&\frac{w_1w_3x_0}{n}+\frac{\lambda w_1^{\alpha}w_3^{\alpha}}{n^{\alpha}}\int_{\frac{w_1w_3x_0}{n}}^{\infty}\frac{1}{t^{\alpha}(1+t)^2}dt\\  \nonumber
&\leq&\frac{w_1w_3x_0}{n}+I[\alpha<1]\frac{\lambda w_1^{\alpha}w_3^{\alpha}}{n^{\alpha}}\int_{0}^{\infty}\frac{1}{t^{\alpha}(1+t)^2}dt+I[\alpha>1]\frac{\lambda w_1^{\alpha}w_3^{\alpha}}{n^{\alpha}}\int_{\frac{w_1w_3x_0}{n}}^{\infty}\frac{1}{t^{\alpha}}dt\\  \label{epp3}
&=& \frac{w_1w_3x_0}{n}+I[\alpha<1]\frac{\lambda w_1^{\alpha}w_3^{\alpha}}{n^{\alpha}}\frac{\alpha\pi}{\sin(\alpha\pi)}+I[\alpha>1]\frac{\lambda x_0^{1-\alpha}}{\alpha-1}\frac{w_1w_3x_0}{n}.
\end{eqnarray}
Note that
\begin{eqnarray}\label{epp4}
\frac{x_0^2}{n^2}\iint_{w_1w_3\leq n,w_1\leq x_0,w_3\leq x_0}f(w_1)f(w_3)w_1^2w_3^2dw_1dw_3&=&O\left(\frac{1}{n^2}\right),
\end{eqnarray}
\begin{eqnarray}\label{epp5}
\frac{1}{n^{2\alpha}}\iint_{w_1w_3\leq n,w_1\leq x_0,w_3\leq x_0}f(w_1)f(w_3)w_1^{2\alpha}w_3^{2\alpha}dw_1dw_3&=&O\left(\frac{1}{n^{2\alpha}}\right),
\end{eqnarray}

\begin{eqnarray}\nonumber
&&\frac{x_0^2}{n^2}\iint_{w_1w_3\leq n,w_1\leq x_0,x_0\leq w_3}f(w_1)f(w_3)w_1^2w_3^2dw_1dw_3\\ \nonumber
&=& \frac{x_0^2}{n^2}\int_{0}^{x_0}f(w_1)w_1^2dw_1\int_{x_0}^{\frac{n}{w_1}}\lambda\alpha w_3^{1-\alpha}dw_3\\ \label{epp6}
&\leq&\frac{x_0^2\lambda\alpha}{n^{\alpha}(2-\alpha)}\int_{0}^{x_0}f(w_1)w_1^{\alpha}dw_1=o\left(\frac{\log n}{n^{\alpha}}\right).
\end{eqnarray}

\begin{eqnarray}\nonumber
&&\frac{1}{n^{2\alpha}}\iint_{w_1w_3\leq n,w_1\leq x_0,x_0\leq w_3}f(w_1)f(w_3)w_1^{2\alpha}w_3^{2\alpha}dw_1dw_3\\ \nonumber
&=& \frac{1}{n^{2\alpha}}\int_{0}^{x_0}f(w_1)w_1^{2\alpha}dw_1\int_{x_0}^{\frac{n}{w_1}}\lambda\alpha w_3^{\alpha-1}dw_3\\ \label{epp7}
&\leq&\frac{\lambda}{n^{\alpha}}\int_{0}^{x_0}f(w_1)w_1^{\alpha}dw_1=o\left(\frac{\log n}{n^{\alpha}}\right).
\end{eqnarray}

\begin{eqnarray}\nonumber
&&\frac{x_0^2}{n^2}\iint_{w_1w_3\leq n,w_1\geq x_0,w_3\geq x_0}f(w_1)f(w_3)w_1^2w_3^2dw_1dw_3\\ \nonumber
&\leq&\frac{x_0^2}{n^2}\lambda^2\alpha^2\int_{x_0}^{n}\int_{x_0}^{\frac{n}{w_1}}w_1^{1-\alpha}w_3^{1-\alpha}dw_1dw_3\\ \nonumber
&=&\frac{x_0^2\lambda^2\alpha^2}{n^{\alpha}(2-\alpha)}\int_{x_0}^{n}w_1^{-1}dw_1\\\label{epp8}
&=&O\left(\frac{\log n}{n^{\alpha}}\right).
\end{eqnarray}

\begin{eqnarray}\nonumber
&&\frac{1}{n^{2\alpha}}\iint_{w_1w_3\leq n,w_1\geq x_0,w_3\geq x_0}f(w_1)f(w_3)w_1^{2\alpha}w_3^{2\alpha}dw_1dw_3\\ \nonumber
&=&(1+o(1))\frac{1}{n^{2\alpha}}\lambda^2\alpha^2\int_{x_0}^{n}\int_{x_0}^{\frac{n}{w_1}}w_1^{\alpha-1}w_3^{\alpha-1}dw_1dw_3\\ \nonumber
&=&(1+o(1))\frac{\lambda^2\alpha}{n^{\alpha}}\int_{x_0}^{n}w_1^{-1}dw_1\\\label{epp9}
&=&(1+o(1))\lambda^2\alpha\frac{\log n}{n^{\alpha}}.
\end{eqnarray}
Hence, by (\ref{epp3})-(\ref{epp9}), we get
\[\mathbb{E}(p_{123}p_{143}I[W_1W_3\leq n])=O\left(\frac{\log n}{n^{\alpha}}\right).\]

Next we show $\mathbb{E}(p_{123}p_{143}I[W_1W_3\leq n])\geq c\frac{\log n}{n^{\alpha}}$ for some constant $c>0$. On the region $w_1w_3\leq n$, $\frac{w_1w_3x_0}{n}\leq x_0$. Then
\begin{eqnarray}\nonumber
\mathbb{E}(p_{123}|W_1=w_1,W_3=w_3)
&\geq&\int_{\frac{w_1w_3x_0}{n}}^{\infty}\mathbb{P}\left(W_2>\frac{nt}{w_1w_3}\right)\frac{1}{(1+t)^2}dt\\  \nonumber
&\geq&\int_{x_0}^{\infty}\mathbb{P}\left(W_2>\frac{nt}{w_1w_3}\right)\frac{1}{(1+t)^2}dt\\
&=&\frac{\lambda w_1^{\alpha}w_3^{\alpha}}{n^{\alpha}}\int_{x_0}^{\infty}\frac{1}{t^{\alpha}(1+t)^2}dt\\  \nonumber
&=&c\frac{w_1^{\alpha}w_3^{\alpha}}{n^{\alpha}},
\end{eqnarray}
for constant $c=\lambda\int_{x_0}^{\infty}\frac{1}{t^{\alpha}(1+t)^2}dt>0$. By (\ref{epp9}), (\ref{epp7}), (\ref{epp5}), we get $\mathbb{E}(p_{123}p_{143}I[W_1W_3\leq n])\geq c\frac{\log n}{n^{\alpha}}$ with a positive constant $c$.

For a constant $C>0$, the variance of $\mathcal{C}_2$ is bounded by
\begin{eqnarray*}
Var(\mathcal{C}_2)
&\leq& \mathbb{E}(\mathcal{C}_2)+C\sum_{\substack{1\leq i_1<j_1<k_1<l_1\leq n\\ 1\leq i_2<j_2<k_2<l_2\leq n\\
1\leq |\{i_1,j_1,k_1,l_1\}\cap\{i_2,j_2,k_2,l_2\}|\leq m-1}}\mathbb{E}\left[p_{i_1j_1k_1}p_{i_1l_1k_1}p_{i_2j_2k_2}p_{i_2l_2k_2}\right]\\
&\leq&\mathbb{E}(\mathcal{C}_2)+C\sum_{\substack{1\leq i_1<j_1<k_1<l_1\leq n\\ 1\leq i_2<j_2<k_2<l_2\leq n\\
|\{i_1,j_1,k_1,l_1\}\cap\{i_2,j_2,k_2,l_2\}|= 1}}\mathbb{E}\left[p_{i_1j_1k_1}p_{i_1l_1k_1}p_{i_2j_2k_2}p_{i_2l_2k_2}\right]\\
&&+C\sum_{\substack{1\leq i_1<j_1<k_1<l_1\leq n\\ 1\leq i_2<j_2<k_2<l_2\leq n\\
2\leq |\{i_1,j_1,k_1,l_1\}\cap\{i_2,j_2,k_2,l_2\}|\leq m-1}}\mathbb{E}\left[p_{i_1j_1k_1}p_{i_1l_1k_1}\right]\\
&\leq&\mathbb{E}(\mathcal{C}_2)+Cn^{7-\alpha}\log n.
\end{eqnarray*}
Hence $\mathcal{C}_2=(1+o_p(1))\mathbb{E}(\mathcal{C}_2)$ if $\alpha<1$ by Markov's inequality.

\subsection{Proof of the case $\alpha>2$ of Theorem \ref{phase2} .}
For $\alpha>2$, the second moment of $W$ exists. Then
\[\mathbb{E}(p_{123}p_{143})\leq \mathbb{E}\left(\frac{W_1W_2W_3}{n}\frac{W_1W_4W_3}{n}\right)=\frac{(\mathbb{E}[W_2])^2(\mathbb{E}[W_1^2])^2}{n^2}.\]
 Hence, $\mathbb{E}[\mathcal{C}_2]=O(n^2)$. Next, we show $\mathbb{E}[\mathcal{C}_2]\geq cn^2$ for a constant $c>0$. Note that for $w_1w_3\leq n$,
 \begin{eqnarray}\nonumber
\mathbb{E}(p_{123}|W_1=w_1,W_3=w_3)
&=&\int_0^{\infty}\mathbb{P}\left(W_2>\frac{nt}{w_1w_3}\right)\frac{1}{(1+t)^2}dt\\ \nonumber
&\geq&\int_{\frac{w_1w_3x_0}{n}}^{x_0^2}\mathbb{P}\left(W_2>\frac{nt}{w_1w_3}\right)\frac{1}{(1+t)^2}dt\\ \label{lowerb}
&\geq&\frac{\lambda}{(1+x_0^2)^2}\frac{w_1^{\alpha}w_2^{\alpha}}{n^{\alpha}}\int_{\frac{w_1w_3x_0}{n}}^{x_0^2}\frac{1}{t^{\alpha}}dt\\ \nonumber
&=&\frac{\lambda}{(1+x_0^2)^2}\frac{w_1^{\alpha}w_2^{\alpha}}{n^{\alpha}}\frac{1}{\alpha-1}\left(n^{\alpha-1}(w_1w_2)^{1-\alpha}x_0^{1-\alpha}-x_0^{2(1-\alpha)}\right)\\ \nonumber
&=&\frac{\lambda}{(1+x_0^2)^2}\frac{x_0^{1-\alpha}}{\alpha-1}\frac{w_1w_2}{n}-\frac{\lambda}{(1+x_0^2)^2}\frac{x_0^{2(1-\alpha)}}{\alpha-1} \frac{w_1^{\alpha}w_2^{\alpha}}{n^{\alpha}}. 
\end{eqnarray}
 
 Hence for a generic positive constant $c>0$,
 \begin{eqnarray}\nonumber
\mathbb{E}(p_{123}p_{143}I[W_1W_3\leq n])&\geq& \frac{c}{n^2}\iint_{w_1w_3\leq n,w_1\geq x_0,w_3\geq x_0}f(w_1)f(w_3)w_1^2w_3^2dw_1dw_3\\ \nonumber
&=&\frac{c}{n^2}\int_{x_0}^{n}\int_{x_0}^{\frac{n}{w_1}}w_1^{1-\alpha}w_3^{1-\alpha}dw_1dw_3\\ \nonumber
&=&\frac{c}{n^{2}(2-\alpha)}\int_{x_0}^{n}w_1^{1-\alpha}(n^{2-\alpha}w_1^{\alpha-2}-x_0^{2-\alpha})dw_1\\ \nonumber
&=&\frac{c}{n^{2}(2-\alpha)}\left(n^{2-\alpha}\log n-\frac{x_0^{2-\alpha}}{2-\alpha}\left(n^{2-\alpha}-x_0^{2-\alpha}\right)\right)\\ \nonumber \nonumber
&=&\frac{1}{n^2}\frac{cx_0^{2(2-\alpha)}}{(2-\alpha)^2}+\frac{c\log n}{n^{\alpha}(2-\alpha)}-\frac{cx_0^{2-\alpha}}{n^{\alpha}(2-\alpha)^2}.
 \end{eqnarray}
 Then $\mathbb{E}[\mathcal{C}_2]\geq cn^2$ for some constant $c>0$ and $\mathbb{E}[\mathcal{C}_2]\asymp n^2$.
 
 Note that
 \[\sum_{\substack{1\leq i<j<k<l\leq n\\1\leq i_1<j_1<k_1<l_1\leq n\\
 |\{i,j,k,l\}\cap\{i_1,j_1,k_1,l_1\}|\geq3}}\mathbb{E}[p_{ijk}p_{ijl}p_{i_1j_1k_1}p_{i_1j_1l_1}]\leq n^5\mathbb{E}[p_{ijk}]=O(n^3).\]

For $\alpha>2$, there exists $\beta$ such that $1<\beta<\min\{2,\frac{\alpha}{2}\}$. Then $\mathbb{E}(W_1^{\beta})$ is finite. For $1\leq i<j<k<l\leq n$ and $1\leq i_1<j_1<k_1<l_1\leq n$,
\[\mathbb{E}(p_{ijk}p_{jkl}p_{i_1j_1k_1}p_{i_1j_1l_1})\leq \sqrt{\mathbb{E}(p_{ijk}^2p_{jkl}^2)\mathbb{E}(p_{i_1j_1k_1}^2p_{i_1j_1l_1}^2)}\leq \sqrt{\mathbb{E}(p_{ijk}^{\beta}p_{ijl}^{\beta})\mathbb{E}(p_{i_1j_1k_1}^{\beta}p_{i_1j_1l_1}^{\beta})}=\frac{O(1)}{n^{2\beta}}.\]
Then
 \[\sum_{\substack{1\leq i<j<k<l\leq n\\1\leq i_1<j_1<k_1<l_1\leq n\\
 |\{i,j,k,l\}\cap\{i_1,j_1,k_1,l_1\}|=2}}\mathbb{E}[p_{ijk}p_{ijl}p_{i_1j_1k_1}p_{i_1j_1l_1}]\leq O(n^{6-2\beta})=o(n^4).\]

For $\alpha>3$, there exists $\beta$ such that $3<2\beta<\min\{4,\alpha\}$. Then $\mathbb{E}(W_1^{2\beta})$ is finite. For $1\leq i<j<k<l\leq n$ and $1\leq i_1<j_1<k_1<l_1\leq n$,
\[\mathbb{E}(p_{ijk}p_{jkl}p_{i_1j_1k_1}p_{i_1j_1l_1})\leq \sqrt{\mathbb{E}(p_{ijk}^2p_{jkl}^2)\mathbb{E}(p_{i_1j_1k_1}^2p_{i_1j_1l_1}^2)}\leq \sqrt{\mathbb{E}(p_{ijk}^{\beta}p_{ijl}^{\beta})\mathbb{E}(p_{i_1j_1k_1}^{\beta}p_{i_1j_1l_1}^{\beta})}=\frac{O(1)}{n^{2\beta}}.\]
Then
 \[\sum_{\substack{1\leq i<j<k<l\leq n\\1\leq i_1<j_1<k_1<l_1\leq n\\
 |\{i,j,k,l\}\cap\{i_1,j_1,k_1,l_1\}|=1}}\mathbb{E}[p_{ijk}p_{ijl}p_{i_1j_1k_1}p_{i_1j_1l_1}]\leq O(n^{7-2\beta})=o(n^4).\]

Hence, for $\alpha>3$, $Var(\mathcal{C}_2)=o\left(n^{4}\right)$.
For any constant $\epsilon>0$, by Markov's inequality, one gets
\[\mathbb{P}\left(\frac{|\mathcal{C}_2-\mathbb{E}(\mathcal{C}_2)|}{n^{2}}>\epsilon\right)\leq\frac{Var(\mathcal{C}_2)}{n^{4}\epsilon^2}=o(1).\]
Then the proof for the case $\alpha>3$ is complete.

\subsection{Proof of the case $\alpha=2$ of Theorem \ref{phase2} .}
When $\alpha=2$, by the proof of the case $\alpha<2$, we have 
\begin{eqnarray}\nonumber
&&\frac{x_0^2}{n^2}\iint_{w_1w_3\leq n,w_1\geq x_0,w_3\geq x_0}f(w_1)f(w_3)w_1^2w_3^2dw_1dw_3\\ \nonumber
&=&\frac{x_0^2}{n^2}\lambda^22^2\int_{x_0}^{n}\int_{x_0}^{\frac{n}{w_1}}w_1^{-1}w_3^{-1}dw_1dw_3\\ \nonumber
&=&2x_0^2\lambda^2\frac{(\log n)^2}{n^2}(1+o(1)).
\end{eqnarray}
Hence $\mathbb{E}(\mathcal{C}_2)\leq cn^2(\log n)^2$ for a constant $c>0$.
By (\ref{lowerb}), we have $\mathbb{E}(\mathcal{C}_2)\geq cn^2(\log n)^2$ for a constant $c>0$. Then the proof is done.

\subsection{Proof of the case $\alpha=1$ of Theorem \ref{phase2} .}
For $\alpha=1$, by  (\ref{epp3}) and (\ref{lowerb}), we have 
\begin{eqnarray}\nonumber
&&\mathbb{E}(p_{123}|W_1=w_1,W_3=w_3)\\ \nonumber
&\leq&\frac{w_1w_3x_0}{n}+\frac{\lambda w_1w_3}{n}\int_{\frac{w_1w_3x_0}{n}}^{x_0^2}\frac{1}{t(1+t)^2}dt+\frac{\lambda w_1w_3}{n}\int_{x_0^2}^{\infty}\frac{1}{t(1+t)^2}dt\\  \nonumber
&\leq&\frac{w_1w_3x_0}{n}+\frac{\lambda w_1w_3}{n}\int_{\frac{w_1w_3x_0}{n}}^{x_0^2}\frac{1}{t}dt+\frac{\lambda w_1w_3}{n}\int_{x_0^2}^{\infty}\frac{1}{t(1+t)^2}dt,
\end{eqnarray}
and for a generic constant $c$,
\begin{eqnarray}\nonumber
\mathbb{E}(p_{123}|W_1=w_1,W_3=w_3)
&\geq&\frac{\lambda}{(1+x_0^2)^2}\frac{w_1w_2}{n}\int_{\frac{w_1w_3x_0}{n}}^{x_0^2}\frac{1}{t}dt=c\frac{w_1w_2}{n}\log\frac{w_1w_2}{n}+c\frac{w_1w_2}{n}. 
\end{eqnarray}
Then by (\ref{epp2}), we have
\begin{eqnarray}\label{alphaone}
\mathbb{E}(p_{123}p_{143}I[W_1W_3\leq n]) 
&=&(1+o(1))\frac{c}{n^2}\iint_{w_1w_3\leq n}\left(\log\frac{w_1w_2}{n}\right)^2dw_1dw_2.
\end{eqnarray}
Straightforward calculation yields
\begin{eqnarray}
\iint_{w_1w_3\leq n}\left(\log w_1\right)^2dw_1dw_2&=&\left(\frac{1}{3}n(\log n)^3-x_0 n(\log n)^2\right)(1+o(1)),\\
\iint_{w_1w_3\leq n}\left(\log n\right)^2dw_1dw_2&=&\left(n(\log n)^3-(x_0+\log x_0) n(\log n)^2\right)(1+o(1)),
\end{eqnarray}
\begin{eqnarray}
\iint_{w_1w_3\leq n}\log w_1\log w_2dw_1dw_2&=&\left(\frac{1}{6}n(\log n)^3-\frac{1}{2}n(\log n)^2\right)(1+o(1)),\\
\iint_{w_1w_3\leq n}\log n\log w_2dw_1dw_2&=&\left(\frac{1}{2}n(\log n)^3-x_0n(\log n)^2\right)(1+o(1)).
\end{eqnarray}
Then by (\ref{alphaone}), $\mathbb{E}(p_{123}p_{143}I[W_1W_3\leq n])=(1+o(1))(x_0-1-\log x_0)c\frac{(\log n)^2}{n}$. We can always take large $x_1>1$ and (\ref{powerlaw}) still holds for $x\geq x_1$ hence the proof still goes through with $x_0$ replaced by $x_1$.
The proof is complete.



\begin{thebibliography}{99}



\bibitem{ABB06}Agarwal, S., K. Branson,  and S. Belongie. 2006. Higher order learning with graphs. 
\textit{Proceedings
of the International Conference on Machine Learning}, 17-24.



\bibitem{ACKZ15}
Angelini, M., F. Caltagirone, F.  Krzakala,  and L. Zdeborova. 2015.
Spectral detection on sparse hypergraphs. 
\textit{Allerton Conference on Communication,
Control, and Computing}, 66-73.



\bibitem{B93} Bolla, M. 1993.
Spectra, euclidean representations and clusterings of hypergraphs.
\textit{Discrete Mathematics},
\textbf{117(1)}, 19-39.



\bibitem{BJR07}
Bollobas, B., Janson, S. and Riordan, O. 2007.
The phase transition in inhomogeneous random graphs.
\textit{Random Structures \& Algorithms}, 31(1): 3-122.

\bibitem{BDM06}
Britton, T. and  Deijfen, M.and Martin-Lof, A.(2006).
Generating simple random graphs with prescribed degree distribution.
\textit{Journal of Statistical Physics}, 124, 1377-1397.


\bibitem{CK10}Chertok, M. and Y. Keller. 2010.
Efficient high order matching. 
\textit{IEEE Trans. on Pattern Analysis and
Machine Intelligence}, \textbf{32(12)}, 2205-2215.

\bibitem{CL06}
Chung, F. and Lu, L.(2006).
The volume of the giant component of a random graph with given expected degrees.
\textit{SIAM J. Discrete Math.}20, 395-411.


\bibitem{CL02}
Chung, F. and Lu, L.(2002).
Connected components in random graphs with given expected degree sequences.
\textit{Annals of Combinatorics}, 6: 125-145.




\bibitem{ER60}Erd\"{o}s, P. and A. R\'{e}nyi. 1960.
 On the evolution of random graphs. \textit{Publ. Math. Inst. Hungar. Acad.
Sci. }, \textbf{5}, 17-61.

\bibitem{ER05} Estrada, E. and J. Rodriguez-velasquez. 2005.
Complex networks as hypergraphs. https://arxiv.org/ftp/physics/papers/0505/0505137.pdf


\bibitem{FO17}
Furedi, Z. and Ozkahya, L. 2017.
On  3-uniform  hypergraphs  without  a  cycle  of  a  givenlength. \textit{Discrete Applied Mathematics}, 216: 582–588.


\bibitem{GD17} Ghoshdastidar, D. and A. Dukkipati. 2017.
Consistency of spectral hypergraph partitioning under planted partition model. \textit{The Annals of Statistics},
\textbf{45(1)}, 289-315.

\bibitem{GL12}
Gyori, E. and Lemons, N. 2012. Hypergraphs with no cycle of length 4. \textit{Discrete  Mathematics}, 312: 1518–1520.



\bibitem{GZCN09}
Ghoshal, G., V. Zlatic, G. Caldarelli, and M. E. J. Newman. 2009. 
Random hypergraphs and their applications. \textit{Physical Review E 79}.



\bibitem{HBF14}
Hu, Z. and Bi, W. and Feng, Q.(2014).
Limit laws in the generalized random graphs with random vertex weights. \textit{Statistics and Probability Letters}, 89:65-76.

\bibitem{HD20}
Hu, Z. and Dong, L.(2020). Number of edges in inhomogeneous random graphs.
\textit{Science China Mathematics}, 63.


\bibitem{HM19}
Huang, H. and Ma, J.(2019).
On tight cycles in hypergraphs.
\textit{SIAM Journal on Discrete Mathematics}. 33(1): 230-237.



\bibitem{J95} Janson, S. (1995).
Random regular graphs: asymptotic distributions and contiguity. 
\textit{Combinatorics, Probability and Computing}, \textbf{4}, 369--405.

\bibitem{JN09}
Jegou, P. and Ndiaye, S.(2009).
On the notion of cycles in hypergraphs.
\textit{Discrete Mathematics}. 309: 6535-6543.


\bibitem{JLS19}
 Janssen, A.J.E.M. and Leeuwaarden, J. and Shneer, S.(2019). Counting cliques and cycles in scale-free inhomogeneous
random graphs.
\textit{Journal of Statistical Physics}, 175:161-184.


\bibitem{JLN10}
Janson, S. and Luczak, T. and  Norros, I.(2010).  Large cliques in a power-law random graph.
\textit{J. Appl. Prob.}, 47:1124-1135.


\bibitem{J07}
Janson, S.(2008).
The largest component in a subcritical random graph with a power law degree distribution. 
\textit{The Annals of Applied Probability}. 18:1651-1668.


\bibitem{KKMO11}
Keevash, P. et al. (2011).
Loose Hamilton cycles in hypergraphs.
\textit{Discrete Mathematics}. 311(7): 544-559.










\bibitem{N01}Newman, M. 2001.
Scientific collaboration networks. I. Network construction and fundamental results.
\textit{Physical Review E}, \textbf{64}, 016-131.



\bibitem{NR06}
Norros, I. and Reittu, H.(2006).
On a conditionally poissonian graph process.
\textit{Adv. Appl. Prob.}. 38:59-75.


\bibitem{QYW14}
Qi,L, Yu, J. and Wang, S.(2014).
Regular uniform hypergraphs, s-cycles, s-paths and their largest Laplacian H-eigenvalues.
\textit{Linear Algebra and its Applications}, 443: 215-227.






\bibitem{SS17}
Solymosi, D. and Solymosi, J.(2017).
Small cores in 3-uniform hypergraphs.
\textit{Journal of Combinatorial Theory, Series B}. 122: 897-910.

\bibitem{S02}
Soderberg, B.(2002)
A general formalism for inhomogeneous random graphs,
\textit{Phys. Rev. E }66, 066121.







\bibitem{YS21}
Yuan, M. and Shang, Z.(2021).
Sharp detection boundaries on testing dense subhypergraph.
https://arxiv.org/abs/2101.04584 






\end{thebibliography}

\end{document}